\documentclass[11pt,twoside]{article}
\usepackage{asp2010}

\resetcounters

\begin{document}

\title{SSOS: A Moving Object Image Search Tool for Asteroid Precovery at the CADC}
\author{Stephen Gwyn$^1$, Norman Hill$^1$, J.J. Kavelaars$^1$}
\affil{$^1$Canadian Astronomical Data Centre,
Herzberg Institute of Astrophysics,
5071 West Saanich Road,
Victoria, British Columbia,
Canada   V9E 2E7}

\begin{abstract}
While regular archive searches can find images at a fixed location,
they cannot find images of moving targets such as asteroids. The Solar
System Object Search (SSOS) at the Canadian Astronomy Data Centre
allows users to search for images of moving objects.  SSOS accepts as
input either a list of observations, an object designation, a set of
orbital elements, or a user-generated ephemeris for an object. It then
searches for observations of that object over a range of dates. The
user is then presented with a list of images containing that object
from a variety of archives. Initially created to search the CFHT
MegaCam archive, SSOS has been extended to other telescope archives
including Gemini, Subaru/SuprimeCam, HST, and several ESO instruments
for a total of 1.6 million images.  The SSOS tool is located on the
web at: \url{http://www.cadc.hia.nrc.gc.ca/ssos}
\end{abstract}

\section{Introduction}
In many fields of astronomy, image archives are of increasing
importance. For example, since 2005, more than 50\% of HST papers
have been based on archival data rather than PI data.  Archival images
have become increasingly useful to extra-galactic and stellar
astronomers in the last few years but, until now, solar system
researchers have been at a disadvantage in this respect.  While
regular archive searches can find images at a fixed location, they
cannot find images of moving targets.

This is unfortunate, because it could be argued that archival data is
potentially more useful to solar system studies than extra-galactic or
stellar astronomy.  The full scientific potential derived from the
discovery of small solar system bodies can not be fully realized until
precise orbital parameters for those objects can be determined.
Often, an object is detected but an orbit can not be determined until
a significant amount of time has passed. If archival (precovery)
images of the object exist, one can determine the orbital parameters
immediately.

The Solar System Object Search (SSOS) at the Canadian Astronomy Data
Centre (CADC) allows users to search for images of moving objects. The user
enters either a set of observations, an object name, orbital elements
or a full ephemeris. SSOS generates an ephemeris and returns a list
of all matching images.

\section{Image Harvesting}
\label{sec:scrape}

Before any queries can be made, the SSOS image database must be
populated. This is done by going to various telescope archives and
harvesting the metadata describing each image taken by the telescope.
For each image SSOS stores the following information: midpoint of
exposure time, RA and Dec of the image center, and the extent of the image
in RA and Dec.  In addition, SSOS stores a bounding box in time, RA
and Dec.  The bounding box is in integer days (MJD) and integer
degrees for RA and Dec.

Obtaining the metadata for images stored at the CADC, where SSOS is
based, is relatively easy. The existing databases describing each
image archive are queried directly and the relevant parameters are
ingested into the SSOS database. To date, the archives of the following
telescopes have been harvested from the CADC: CFHT (MegaCam and
WIRCam), Gemini (GMOS), and HST (WFPC, ACS, and WFC3).

Offsite archives must be ``scraped'' over the web. This can take many
forms depending on the archive. The Subaru SuprimeCam image lists are
available as simple ASCII text files. The ESO archive can be queried
repeatedly by date, eventually returning all the observations made by
those telescopes. To date, the archives of the following telescopes
have been harvested: 
AAT (WFI),
ESO-LaSilla 2.2m (WFI),
ESO-NTT (EFOSC, EMMI, SOFI, SUSI/SUSI2),
ESO-VISTA (VIRCAM),
VLT (FORS1, FORS2, HAWKI, NAOS-CONICA, ISAAC and VIMOS), and
Subaru (SuprimeCam).

\begin{figure}
\plotone{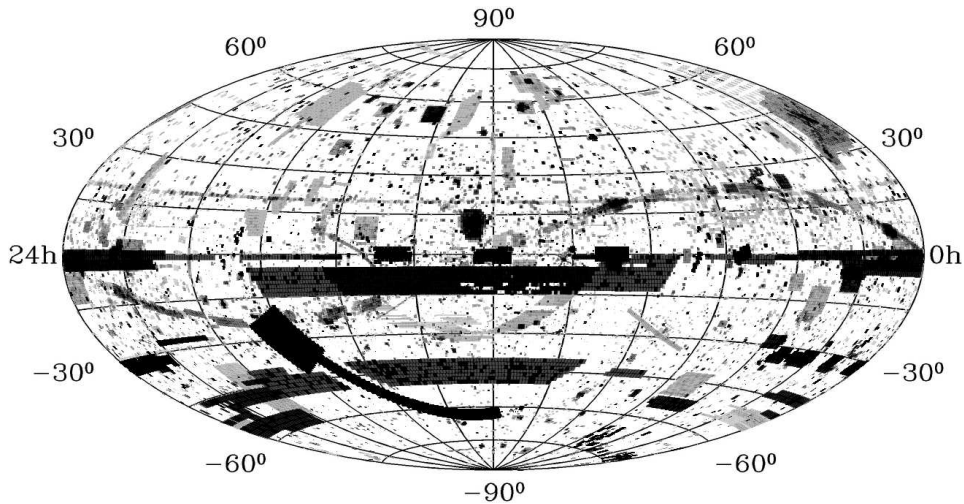}
\caption{Area of the sky covered by SSOS. The greyscale gives an
  indication of the number of images covering a particular spot on the
  sky, with a single image being represented by the faintest grey, and
  40 or more images being indicated by solid black.}
\label{fig:coverage}
\end{figure}

Currently, there are 1.6 million images in the SSOS database.  Figure
\ref{fig:coverage} shows the area of the sky covered by SSOS.

\section{User Input and Conversion to Ephemeris}
\label{sec:input}

When arriving at the Solar System Object Search tool
website, users have four
ways to search for images. In each case, SSOS converts the user's
input into an ephemeris. The four methods of input and conversion are
detailed in the following four subsections:

\subsection{Search by arc}
In this input method, the user enters a series of observations in MPC
format.  SSOS then uses these observations to determine an orbit and
generate an ephemeris from that orbit.  The user can select one of two
orbit fitting routines: The orbit fitting code of \citet{bernstein}
has been set up to automatically convert the observations into orbital
parameters ({\tt fit\underline{~}radec}) and use those parameters to
produce an ephemeris ({\tt predict}). SSOS also provides the new object
ephemeris generator from the Minor Planet
Center\footnote{\url{http://www.minorplanetcenter.net/iau/MPEph/NewObjEphems.html}}
as an alternative.  If a user selects this option, the SSOS queries
the MPC service automatically.  The MPC fits a V\"ais\"al\"a orbit to the
observations and returns an ephemeris based on this orbit.  This
method is slower than the Bernstein \& Khushalani fitting because it
requires SSOS to make queries to an external service.  The ephemeris is
generated at intervals of 24 hours.  Mauna Kea (observatory code 568)
is used as the observing site.

\subsection{Search by object name}
In this input method, the user enters the name of an object.  SSOS
then forwards that name to one of two services, either the Lowell
Observatory asteroid ephemeris
generator\footnote{\url{http://asteroid.lowell.edu/cgi-bin/asteph}} or the
minor planet and comet ephemeris service at the Minor Planet
Center\footnote{\url{http://www.minorplanetcenter.net/iau/MPEph/MPEph.html}}. These
services query their databases for an object matching the name, make
the appropriate orbital calculations and return an ephemeris to SSOS.

In addition to using these two offsite services, SSOS can also
generate an ephemeris locally. SSOS maintains a regularly updated copy
of the MPC orbital element
database\footnote{\url{http://www.minorplanetcenter.net/iau/MPCORB.html}}.
When a user enters an object name, the local version of this database
is queried and the orbital elements are passed to the program {\tt
orbfit} from the OrbFit software
package \citep{orbfit}\footnote{\url{http://adams.dm.unipi.it/~orbmaint/orbfit/}},
which generates an ephemeris.
As with the search by arc option, the ephemeris is
generated at 24 hour intervals and Mauna Kea is used as the observing
site.

\subsection{Search by orbital elements}
In this case, the user enters the orbital elements of an object: epoch,
semi-major axis, eccentricity, inclination, longitude of the ascending
node, argument of perihelion and mean anomaly.  These orbital elements
are used as input to the program {\tt orbfit} which returns an
ephemeris, again at 24 hour intervals and using Mauna Kea as the
observing location.

\subsection{Search by ephemeris}
\label{ssec:ephem}
This method allows the user complete control over the ephemeris. The
user enters a series of times and object positions. Users can ``cut
and paste'' text into the service.  This method is useful if the user
has any concerns about the positional accuracy of any of the previous
methods. For example, the object in question might be near enough to
the earth that the parallax will significantly affect the objects
positions. Alternatively, the object's apparent motion might be
irregular enough that the linear daily interpolation scheme is
insufficiently accurate.

\section{Searching along the ephemeris}
\label{sec:match}

Once an ephemeris has been generated by one of the above methods, the next
step is to match that ephemeris to the database of observations.  The
ephemeris can be thought of as a series of line segments in 3
dimensions: time, RA and Dec. SSOS builds an integer-valued bounding
box (in both time and position) around each segment. The ephemeris is
saved as a temporary table with both the integer bounding boxes and
the exact (floating point) values.  When the object moves across the
first point of Aries, two rows are added, one each describing the
position of the object on either side of the celestial meridian.
 
SSOS then matches this temporary table to its observation table. For
speed, the integer bounding boxes are matched first. If a match is
possible ({\it i.e.}, if the bounding boxes overlap) SSOS does a linear
interpolation to determine the position of the object at the time of
the exposure. A match occurs if this position is within the footprint
of the image. 

The key to keeping the queries reasonably fast are the integer
bounding boxes and the linear interpolation. Doing a full orbital
prediction for each of the images is not feasible. This is
sufficiently accurate for the majority of queries, where the object
either moves slowly or in a fairly straight line.  A typical 20 year
ephemeris can be matched to 1.6 million images in less than a second.

SSOS then returns a list of matching images to the user. Where
possible, direct links to the images are provided; otherwise, SSOS
provides links to pages where the images can be requested.  On
average, a search will return 20 images of a given asteroid, but this
number can range from 0 (no hits) to several hundred.


\end{document}